\documentstyle[eqsecnum,aps,preprint]{revtex}

\begin{document}
\draft
\title{Relativistic versus nonrelativistic optical potentials in
$A(e,e'p)B$ reactions}

\author{
J. M. Ud\'\i as$^{(1)}$, P. Sarriguren$^{(2)}$, E. Moya de
Guerra$^{(2)}$, E. Garrido$^{(2)}$, and J. A. Caballero$^{(2)}$
}
\address{
$(1)$ National Laboratory for Nuclear and High--Energy Physics,
Section K (NIKHEF--K), \\
P.O. Box 41882, 1009 DB Amsterdam, The Netherlands\\
$(2)$
Instituto de Estructura de la Materia,
Consejo Superior de Investigaciones Cient\'\i ficas, \\
Serrano 119, E--28006 Madrid, Spain\\
}
\maketitle
\begin{abstract}

We investigate the role of relativistic and nonrelativistic optical
potentials  used in the analysis of ($e,e'p$) data. We find that
the relativistic calculations produce smaller ($e,e'p$) cross
sections even in the case in which both relativistic and
nonrelativistic optical potentials fit equally well the elastic
proton--nucleus scattering data. Compared to the nonrelativistic
impulse approximation, this effect is due to a depletion in the
nuclear interior of the relativistic nucleon current, which should
be taken into account in the nonrelativistic treatment by a proper
redefinition of the effective current operator.

\end{abstract}

\section{Introduction}
\label{sec:intro}

The quasielastic $(e,e'p)$ reaction has been extensively studied
over the last years as a powerful tool to obtain information on the
momentum distribution of the nuclear bound states and to extract
experimental information on absolute spectroscopic factors.

Although high precision measurements of cross sections for this
reaction are already available \cite{deW90,Quint,Kra89}, the
extraction of spectroscopic factors from experiment is still not
free of ambiguities. The origin of the uncertainty has to be found
in the complexity of the reaction and the different approaches
proposed to handle it, that produce different cross sections even
within the impulse approximation (IA) scheme considered here.
It is clear that a reliable determination of spectroscopic factors
requires an accurate description of the mechanism of the reaction.

One major puzzle at present is the discrepancy between
spectroscopic factors obtained from relativistic and nonrelativistic
analyses of data.

Traditionally, differential cross sections for quasielastic
electron--nucleus scattering have been calculated using
nonrelativistic approaches to the nuclear currents. The analyses of
($e,e'p$) data are generally made (see references
\cite{deW90,Quint,Kra89} and references therein) within this
nonrelativistic framework using  the {\sc dweepy} program
\cite{GP,GP4}, which provides a rather complete description of the
process. A fully relativistic formalism for the quasielastic
($e,e'p$) reaction has appeared over the last years
\cite{Pick85,Pick87,Pick89}, and applications to the extraction of
spectroscopic factors comparing with the experimental data have
become available recently \cite{Joe,Jin,Udietal}.

The typical values of the spectroscopic factors obtained within
these relativistic analyses (about 70\% for the $3s_{1/2}$ state
in $^{208}$Pb) are much larger than those obtained in the
nonrelativistic analyses (about 50\% for the same shell as above
\cite{Quint}). These higher values are consistent with theoretical
predictions \cite{theor} as well as with the spectroscopic factors
obtained from other methods \cite{Wagner}. Yet, the difference with
respect to the nonrelativistic results is distressing and remains
to be explained.

In Ref. \cite{Udietal} we studied the differences between the
relativistic and the nonrelativistic treatments of the ($e,e'p$)
reaction and, in particular, we investigated the causes leading to
the discrepancies found in the spectroscopic factors obtained in
the two formalisms. Two different aspects of the analysis were
identified in said reference as the main sources of discrepancy.
First, the treatment of the Coulomb distortion of the electron,
which at present is only exact within a relativistic formalism.
We demonstrated that a complete treatment of this distortion is
necessary in order to obtain reliable spectroscopic factors in
heavy nuclei. Second, the different quenching of the ($e,e'p$)
cross section produced by the relativistic and the nonrelativistic
optical potentials, which are introduced to take into account final
state interactions. We showed that this quenching can differ
typically by 15\%, even though both relativistic and nonrelativistic
optical potentials fit the  elastic proton--nucleus scattering data
for the particular proton energies and mass target nuclei under
study. In this paper we elaborate more on this last point.

The optical potentials used in ($e,e'p$) are generally determined
from elastic nucleon--nucleus scattering data. It is well known
that these data are only sensitive to the asymptotic  behavior of
the wave functions. Wave functions that are different in the nuclear
interior but are identical in the asymptotic region give rise to
equal elastic observables. However, this is not necessarily the case
for inelastic ($p,p'$) scattering or  for ($e,e'p$) reactions.

In Ref. \cite{Sher86} it was shown that the results for inelastic
($p,p'$) scattering may differ when using different optical
potentials that give nearly equivalent fits to the elastic
observables. In particular, in that reference, results obtained with
a Dirac--equation--based (DEB) optical potential were presented. As
discussed in the next section, the DEB potential is obtained from
the relativistic optical potential when the Dirac equation is
transformed into a Schroedinger--like equation for the upper
component. Though derived in this particular way the DEB potential
can be used in the nonrelativistic formalism as another
phenomenological optical potential. The advantage of treating the
DEB potential on the same footing as other nonrelativistic optical
potentials is  that with this potential the Schroedinger equation
produces the same elastic scattering as the Dirac equation.

Concerning the ($e,e'p$) process, several questions have emerged
in the last years, namely, (i) to what extent different optical
potentials fitted to elastic proton--nucleus scattering data may
differ in their predictions on $(e,e'p)$ cross sections?, (ii) what
are the features of these optical potentials to which the $(e,e'p)$
reaction is sensitive while elastic proton--nucleus scattering is
not?, (iii) is the nuclear interior probed by the ($e,e'p$) reaction
responsible for the discrepancies found between the relativistic
and the nonrelativistic approaches?. In this context, some properties
of final state interactions and optical potentials have been already
studied within a nonrelativistic framework. In Ref. \cite{Boffi},
the role of nonlocality in the treatment of final state interactions
and its effect on the extracted occupation numbers from $(e,e'p)$
was investigated, and the estimated effect was about a 15\% increase
in the occupation probabilities. In Ref. \cite{Blok87} a
phenomenological analysis was carried out to show that the ($e,e'p$)
cross sections are sensitive  to the behavior of the optical
potential in the nuclear interior. In this last reference it was
also  argued that an increased absorption in the nuclear interior,
with respect to the absorption produced by the traditional
parametrization of the optical potentials, is more consistent both
with ($e,e'p$) data and with microscopic calculations of the optical
potentials. These arguments were already taken into account in
constructing the nonrelativistic optical potentials given in Refs.
\cite{Quint,Kra89} and used in this work under the name standard
nonrelativistic optical potentials.

In this paper we compare the ($e,e'p$) differential cross section
obtained with the nonrelativistic treatment provided by the
{\sc dweepy} program using  different nonrelativistic optical
potentials,  as well as with the results obtained with the fully
relativistic treatment \cite{Udietal}. We show that the results
for ($e,e'p$) with relativistic and nonrelativistic optical
potentials differ even in the case in which both types of potentials
give exactly the same results for elastic proton--nucleus scattering.
We also explore the reasons for the discrepancies.

The paper is organized as follows: In Sec. II we discuss the choices
taken for the optical potentials within the relativistic and the
nonrelativistic formalisms, and what are their distinguishing
features, focusing on $^{208}$Pb. In Sec. III we summarize briefly
the relativistic and nonrelativistic formalisms used in this work,
and discuss the results for $(e,e'p)$ cross sections  for $^{208}$Pb
obtained with various potentials. Some results for $^{40}$Ca are
also given.  In Sec. IV we present the main conclusions.

\section{Relativistic and nonrelativistic optical potentials}

The usual procedure to take into account final state interactions
in the $(e,e'p)$ reaction is to introduce as input a
phenomenological optical potential with parameters fitted to
reproduce elastic proton--nucleus scattering data. Two different
approaches are widely used in the construction of the optical
potentials, which correspond to relativistic or nonrelativistic
descriptions of the proton--nucleus scattering. Though
microscopically derived optical potentials are available in the
literature (for a recent review see Ref. \cite{Ray92}), in this work
we use empirical parametrizations. As in our previous work
\cite{Udietal}, we use phenomenological optical potentials based on
complex central and spin--orbit potentials, in the nonrelativistic
case, and on standard Lorentz scalar and time--like vector complex
terms (S--V) in the Dirac phenomenology.

To be specific, in the relativistic case we use the parametrization
denoted as fit 2 in Ref. \cite{Hama} of the scalar $(U_{\rm S})$,
vector $(U_{\rm V})$ and Coulomb $(U_{\rm C})$ potentials to solve
the time independent Dirac equation in configuration space:
\begin{equation}
[ i \mbox{\boldmath $\alpha \cdot \nabla $}-\beta (M+U_{\rm S}) +
E - U_{\rm V} - U_{\rm C} ] \Psi = 0 \; ,
\label{dirac}
\end{equation}
where $\Psi\equiv (\Psi_{\rm up},\Psi_{\rm down}) $ is a Dirac
four--spinor. The potentials of Ref. \cite{Hama}  are obtained from
global fits whose parameters are functions of both projectile
energy and target mass number. The parameters have been fitted to
elastic proton--nucleus observables (cross sections, analyzing
powers, and spin rotation functions) and the range of applicability
covers spherical nuclei with mass numbers 40$<A<$208, and energies
65 MeV $<E<$ 1040 MeV. A new parametrization has been reported
recently \cite{Coop93}, extending the range of applicability to 20
MeV and including $^{12}$C and $^{16}$O in the fit. For the mass
number and proton energy of our concern here, the agreement with
experiment is comparable to that of Ref. \cite{Hama}.

In the nonrelativistic treatment we use for the outgoing proton the
solutions of the Schroedinger equation with two types of potentials:
{\em i)} the DEB potential, obtained from the relativistic one as
discussed below, and {\em ii)} the phenomenological parametrization
of Ref. \cite{Quint}, involving Coulomb, complex central, imaginary
surface and complex spin--orbit terms:

\begin{eqnarray}
U(r)&=&-V_{\rm C} (r,R_{\rm C})-V f(x_{\rm V})-i W
 f (x_{\rm W})+ 4 i a_{\rm S} W_{\rm S} f(x_{\rm S})
\nonumber \\
&& +(2/r) (V_{\rm SO} f'(x_{\rm SO})+i W_{\rm SO} f'(x_{\rm WSO}))
\mbox{\boldmath $\sigma \cdot l $}
\label{nrpot}
\end{eqnarray}
where $x_i=(r-R_i)/a_i$ ($i={\rm V,W,S,SO,WSO,C})$,
$R_i=r_i (A-1)^{1/3}$, $f'(x)=df(x)/dx$, $f(x)$ is the standard
Woods--Saxon function, $V_{\rm C}(r,R_{\rm C})$ is the Coulomb
potential of a homogeneously charged sphere with radius
$R_{\rm C}=\sqrt{5/3}<\! r^2\! >^{1/2}$. The parameters are given
in table I.

The DEB potential is obtained by rewriting the Dirac equation
(Eq.(\ref{dirac})) as a second order differential equation for the
upper component (see Appendix A) to obtain the equivalent
Schroedinger equation:
\begin{equation}
\left[ - \frac{ \mbox{\boldmath $\nabla$}^2  }{2 M} - U_{\rm DEB}
\right] \phi({\bf r})= E_{\rm nr} \phi({\bf r}) \; ,
\label{schr}
\end{equation}
with $E_{\rm nr}=(E^2-M^2)/2M$ and $\phi({\bf r})$ a bi--spinor.
This is the standard procedure \cite{Sher86} used to analyze the
relationship between the large S--V potentials in the Dirac
phenomenology, and the usual spin independent and spin--orbit
potentials in the Schroedinger equation. As it is known
\cite{Sher86}, the DEB potential contains an effective central
potential that results from a partial cancellation between the
S--V relativistic potentials plus important quadratic terms, and
a spin--orbit potential that originates from additive contributions
from the S--V potentials:
\begin{equation}
U_{\rm DEB}= V_{\rm C} + V_{\rm so}
\mbox{\boldmath $\sigma \cdot l $}
\label{udeb}
\end{equation}
where
\begin{eqnarray}
V_{\rm C}&=&\frac{1}{2M} [ (U_{\rm V}+U_{\rm C})^2-2E(U_{\rm V}+
U_{\rm C})-U_{\rm S}^2 \nonumber \\
&&-2MU_{\rm S} + V_{\rm D}] \; ,
\label{vc}
\end{eqnarray}
with
\begin{equation}
V_{\rm D}=\frac{1}{rA}\frac{\partial A}{\partial r}+
\frac{1}{2A} \frac{\partial ^2A}{\partial r^2}-\frac{3}{4A^2}
\left( \frac{\partial A}{\partial r}\right) ^2\; ,
\label{vd}
\end{equation}
\begin{equation}
A(r)=\frac{E-U_{\rm V}-U_{\rm C}+M+U_{\rm S}}{E+M}\; ,
\label{ar}
\end{equation}
and
\begin{equation}
V_{\rm so}=\frac{1}{2M} \frac{1}{rA} \frac{\partial A}{\partial r}
\; .
\label{vso}
\end{equation}

A well known feature of this procedure \cite{Ray92} is that the DEB
potentials, and in particular the real central part, show a more
dramatic energy dependence than the standard potentials with
Woods--Saxon shapes. This is specially important for proton energies
larger than $\sim $200 MeV where the real central DEB potential
weakens its attraction in the interior of the nucleus but not at the
surface \cite{Ray92}. The departure from standard Woods--Saxon
shapes is characteristic of the Dirac approach, and is due to the
presence of nonlinear terms in the central potential. Even when the
S--V potentials have standard Woods--Saxon shapes, the
nonrelativistic potentials obtained from them will in general
have nonstandard geometries. Although these changes start to be
sizable above proton energies of 200 MeV, they are also present
to a lower extent at the energies of interest in this work
(100 MeV for the ejected proton in the $(e,e'p)$ reaction).

Figure 1 contains the real and imaginary parts of the DEB potential,
for the particular case of $^{208}$Pb and for a proton energy of
100 MeV, compared with the  standard nonrelativistic optical
potential given by Eq. (\ref{nrpot}) and table I. The DEB potential
has been obtained using Eqs. (\ref{vc}--\ref{vso}) and the
phenomenological S--V  relativistic potential of fit 2 in Ref.
\cite{Hama}, calculating the Coulomb potential $U_{\rm C}$ from the
empirical charge distribution of the target nucleus, as it is done
in said reference. As can be seen from Fig. 1 the real central
potentials, which include the Coulomb contributions, are similar
in shape at this energy, DEB being deeper in the interior and more
repulsive at the surface. The depth of the imaginary central
potential is also larger for DEB showing a departure from a
Woods--Saxon shape near the surface. We have checked that these
features prevail independently on whether we use fit 1 or fit 2
of Ref. \cite{Hama}, or even the EDAI fit of Ref. \cite{Coop93},
for the nucleus and energy considered here. Actually these three
fits produce very similar fully relativistic $(e,e'p)$ cross
sections \cite{Udi93}. It should be mentioned, however, that the
imaginary part of the DEB central potential varies for these
three fits, and depends more than the real part on the particular
choice of the S--V relativistic potential. For the three fits
above mentioned, the real parts of $V_C$ are practically identical,
while the imaginary parts have somewhat different depths. We
consider here fit 2 because the imaginary part of $V_{\rm C}$
produced by fit 2 is the shallowest and closest to the standard
potential.

The spin--orbit potentials show a similar shape in both the DEB and
standard potentials, with a somewhat larger strength for DEB.

Clearly the comparison in Fig. 1 is useful to understand the
relationship between results obtained in the nonrelativistic
treatment of $(e,e'p)$ with DEB and standard optical potentials,
as discussed in next section. In addition, using the DEB potential
helps to understand the relationship between results of relativistic
and nonrelativistic treatments.

It can be shown (see Appendix A) that the solution of the equivalent
Schroedinger equation (Eq. (\ref{schr})) is related to the upper
component of the solution of the Dirac equation (Eq. (\ref{dirac}))
by:
\begin{equation}
\Psi_{\rm up}({\bf r}) =K(r) \phi({\bf r})
\label{psiup}
\end{equation}
with
\begin{equation}
 K(r)=A^{1/2}(r)\; .
\label{kr}
\end{equation}
In Fig. 2 we show the real and imaginary parts of $A(r)$ as obtained
from  Eq. (\ref{ar}). The imaginary part is very small and has been
neglected in the subsequent calculations. The real part is clearly
different from unity in the interior of the nucleus. We get a steady
value of about 0.63 ($K(r)\sim 0.79$) in the interior going to unity
asymptotically.

Thus the distorted wave $\phi({\bf r})$ generated by the DEB
potential is equal to the upper component of the Dirac equation only
asymptotically ($\lim_{r\rightarrow \infty} K(r)=1$). This means
that Eq. (\ref{dirac}) and Eq. (\ref{schr}) will produce the same
elastic proton--nucleus observables, which are only sensitive to
this asymptotic behavior, but for processes where the nuclear
interior plays a role, both equations can lead to different results.
On the other hand, solving the Schroedinger equation (\ref{schr})
for $\phi({\bf r})$ and using the Darwin factor $K(r)$ is equivalent
to solving the Dirac equation (\ref{dirac}) for
$\Psi_{\rm up}({\bf r})$. Hence, comparing results for $(e,e'p)$
with relativistic and nonrelativistic treatments based on the same
relativistic potential, allows us to disentangle effects due to
different features of the optical potentials from effects due to
the fully relativistic treatment.

\section{Results for $\lowercase{(e,e'p)}$ cross sections}

In this section we first summarize briefly the formalism used to
describe the ($e,e'p$) reaction both relativistically and
nonrelativistically. More details can be found in Refs.
\cite{Jin,Udietal,Udi93}. We base our calculations on the impulse
approximation (virtual photon absorbed by the detected nucleon),
which is known \cite{Fru84} to be a reliable approximation at
quasielastic kinematics.

In Refs. \cite{Joe,Jin,Udietal,Udi93} it has been shown the
importance of treating correctly the electron Coulomb distortion,
specially in heavy targets, in order to obtain reliable
spectroscopic factors from experiment. For the purpose of this
work it is however advantageous to switch off the electron
Coulomb distortion, treating the electron current in plane wave
Born approximation (PWBA). The reasons for this are that the
role of the various optical potentials stands out more clearly
and that at present the electron Coulomb distortion cannot be
treated exactly within the nonrelativistic framework. Hence, in
this work all the calculations are made in PWBA (no electron
Coulomb distortion) and the differences in the results presented
come only from the various approximations to the nuclear current.
In impulse approximation, differences between relativistic and
nonrelativistic analyses can occur due to the bound nucleon wave
function, to the current operator, or to the final nucleon wave
function in the $\gamma NN$ vertex. Therefore, we first discuss
the choice of these ingredients within the two formalisms.

All the results in this section correspond to ($e,e'p$) reduced
cross sections in parallel kinematics (momentum transfer parallel
to missing momentum, ${\bf q} \parallel {\bf p}$) with a fixed
value of the kinetic energy of the outgoing proton ($T_F=$100 MeV).
Since these results do not include electron Coulomb distortion, we
do not compare them with experiment (for such a comparison see
Ref. \cite{Udietal}).

\subsection{Relativistic formalism}

Results for the $(e,e'p)$ reaction obtained through a fully
relativistic formalism have appeared in the last years, either
computing the nuclear matrix elements in configuration space
\cite{Joe,Jin,Udietal} or in momentum space
\cite{Pick85,Pick87,Pick89}. While the latter formalism may be
somewhat more elegant and better suited to deal with $p$--dependent
terms in the current operator, the first one is more adequate when
the Coulomb distortion of the electron wave functions has to be
taken into account.

For the relativistic formalism in configuration space, that we use
here, the basic equations that determine the reduced cross section
are given explicitly in Refs. \cite{Jin,Udietal}, in terms of the
electron and nuclear currents. The calculations have been performed
with the relativistic code developed by one of us \cite{Udi93}.
We give here the basic equations in PWBA.

We work in the laboratory frame in which the target nucleus is at
rest and use the notation and conventions of Ref. \cite{BD64}. We
denote by $k_i^\mu=(\epsilon_i,{\bf k}_i)$ the four--momentum of the
incoming electron and by $k_f^\mu=(\epsilon_f,{\bf k}_f)$ the
four--momentum of the outgoing one. The four--momentum of the
exchanged photon is $q^\mu=k_i^\mu-k_f^\mu=(\omega,{\bf q})$.
$P_A^\mu=(M_A,{\bf 0})$ and $P_{A-1}^\mu =(E_{A-1}, {\bf P}_{A-1})$
denote the four--momenta of the target and residual nucleus, while
$P_F^\mu=(E_F,{\bf P}_F)$ is the four--momentum of the ejected
proton.

Using plane waves for the electrons and considering knock--out from
a given $\{nlj\}$ shell, we write the amplitude for the $(e,e'p)$
process in DWIA as \cite{Udietal,Udi93,Fru84}:

\begin{equation}
W_{if}=\frac{m_e}{\sqrt{\epsilon_i \epsilon_f}}
\bar{u}({\bf k}_f,\sigma_f)\gamma_\mu u({\bf k}_i,\sigma_i)
\frac{(-1)}{q_\mu^2}
\sqrt{N_{\{nlj\}}} J_N^\mu (\omega,{\bf q})\; ,
\label{wif}
\end{equation}
where $u({\bf k},\sigma)$ represent four--component relativistic
free electron spinors \cite{BD64}, $N_{\{nlj\}}$ is the occupation
number of the $\{nlj\}$ shell, and $J_N^\mu (\omega,{\bf q})$ is
the nucleon current
\begin{equation}
J_N^\mu(\omega,{\bf q})=
\int\!\!\!  d{\bf y} e^{i{\bf q}\cdot {\bf y}}
\bar{\Psi}_F ({\bf y}) \hat{J}^\mu _N \Psi_B({\bf y}) \; ,
\label{jmucurrent}
\end{equation}
where $\Psi_B$ and $\Psi_F$ are the wave functions for the initial
bound nucleon and for the outgoing final nucleon, respectively, and
$\hat{J}^\mu_N$ is the nucleon current operator to be specified
later. These are the three ingredients that change when one
considers a relativistic treatment or a nonrelativistic one.

Within the relativistic framework the bound state wave function
for the proton, $\Psi_B$, is a four--spinor with well defined
angular momentum quantum numbers $\kappa_B \mu_B$ corresponding to
the shell under consideration. We use four--spinors of the form

\begin{equation}
\Psi^\mu_\kappa ({\bf r})=\left(\begin{array}
{@{\hspace{0pt}}c@{\hspace{0pt}}}
g_{\kappa}(r)
\phi^\mu_\kappa(\hat{r}) \\
 i f_{\kappa}(r) \phi^\mu_{-\kappa}(\hat{r})\end{array}\right)
\label{psimuk}
\end{equation}
that are eigenstates of total angular momentum with eigenvalue
$j=|\kappa|-1/2$,
\begin{equation}
\phi^{\mu}_{\kappa}(\hat{r}) = \sum_{m,\sigma}
< l \ m \ \frac{1}{2} \ \sigma | \ j \ \mu >  Y_{lm}(\hat{r})
\chi_{\sigma}^{\frac{1}{2}}
\label{fimuk}
\end{equation}
with $l=\kappa$ for $\kappa > 0 $, $l=-\kappa -1$ for $\kappa < 0$.
$f_\kappa$ and $g_\kappa$ are the solutions of the usual radial
equations \cite{Rose}. The mean field in the Dirac equation is
determined through a Hartree procedure from a relativistic
lagrangian with scalar and vector S--V terms \cite{SW86}. We use the
parameters of Ref. \cite{HS}, and the {\sc timora} code
\cite{HSbook}.

The wave function for the outgoing proton $\Psi_F$, is a scattering
solution of the Dirac equation (\ref{dirac}), which includes S--V
global optical potentials, as discussed in Sec. II.  This wave
function is obtained as a partial wave expansion in configuration
space

\begin{eqnarray}
\Psi_F({\bf r})=4 \pi \sqrt{\frac{E_F+M}{2E_F}}
&{\displaystyle \sum_{\kappa,\mu,m}}&
 e^{-i\delta^{*}_{\kappa}}
i^{l} < l \ m \ \frac{1}{2} \ \sigma_F | j \ \mu >\nonumber \\
&& \times Y_{lm}^{*}(\hat{P}_F)\Psi_{\kappa}^{\mu}({\bf r}) \; ,
\label{expan}
\end{eqnarray}
where $\Psi_{\kappa}^{\mu}({\bf r})$ are four--spinors of the same
form as that in Eq. (\ref{psimuk}), except that now the radial
functions $f_\kappa$, $g_\kappa$ are complex because of the complex
potential. It should also be mentioned that since the wave function
(\ref{expan}) corresponds to an outgoing proton, we use the complex
conjugates of the radial functions and phase shifts (the latter with
the minus sign).

For the nucleon current operator we take the free nucleon expression
\begin{equation}
\hat{J}_{N}^\mu =F_1\gamma^\mu+i \frac{\bar{\kappa}F_2}{2M}
\sigma^{\mu\nu}q_\nu \; ,
\label{cc2}
\end{equation}
where $F_1$ and  $F_2$ are the nucleon form factors related in the
usual way \cite{BD64} to the electric and magnetic Sachs form
factors of the dipole form.  As discussed in Refs.
\cite{Udietal,Chin92}, DWIA results depend on the choice of the
nucleon current operator. Here we have chosen the operator that is
closer to the one used in the nonrelativistic calculations in the
{\sc dweepy} code.

The numerical calculations involve: i) computation of the radial
functions in configuration space; ii) numerical integration; and
iii) summation of partial waves. The accuracy of the numerical
procedure is tested by comparison to the exact result in the plane
wave limit (PWIA). Typically our calculations involve 30--40
partial waves for the ejected proton and numerical integration
over a range of 15--20 fm in steps of 0.1 fm. The parameters
(number of partial waves, radii and step size of integration)
are adjusted so that differences with exact PWIA cross sections are
less than 0.1\%. Finally, it should also be mentioned that recoil
effects, though small, are taken into account by replacing ${\bf q}$
in Eq. (\ref{jmucurrent}) by ${\bf q} (A-1)/A$, and ${\bf P}_F$ by
$({\bf P}_F (A-1)+{\bf P}_{A-1})/A$.

\subsection{Nonrelativistic formalism}

In the nonrelativistic formalism, the numerical calculations have
been done with the code {\sc dweepy} \cite{GP} that uses as input
nonrelativistic optical potentials and bi--spinor bound nucleon
wave functions. As already indicated, the calculations have been
done switching off the Coulomb distortion of the electron wave
functions. The nucleon current operator used in this code was
obtained \cite{GP4,McV} through a Foldy--Wouthuysen procedure
up to order $(p/M)^4$, based on the current operator in Eq.
(\ref{cc2}) for free nucleons satisfying the relation
\begin{equation}
\Psi_{\rm down}=\frac{\mbox{\boldmath $\sigma$}\cdot {\bf p}}{E+M}
\Psi_{\rm up}
\label{downup}
\end{equation}

Thus, at variance with the relativistic formalism, the nucleon
current in Eq. (\ref{jmucurrent}) is calculated using a
nonrelativistic current operator  and bi--spinors for the initial
(bound) and final (scattering) proton wave functions. The outgoing
proton wave functions are obtained as solutions of the Schroedinger
equation with the DEB and standard optical potentials defined in
Sec. II. The bound proton wave functions are generally obtained as
solutions of the Schroedinger equation with real central and
spin--orbit Woods--Saxon type potentials \cite{Udietal}. However,
for the nonrelativistic results presented here we have used instead
the upper component of the relativistic bound nucleon wave function,
normalized to one. This minimizes differences in the cross sections
coming from the use of different bound nucleon wave functions in
the relativistic and nonrelativistic formalisms. Indeed, we have
checked that in this case both formalisms give the same results in
PWIA  (i.e., in the limit of no final state interactions). Thus,
with this choice, we ensure that the differences among the various
results presented here are solely due to differences in the outgoing
nucleon wave functions generated from the different optical
potentials. Though in principle in the nonrelativistic case one
would take the upper component of the relativistic bound state
wave function after projection of the positive energy part,
in practice for the solutions of the {\sc timora} code this makes
practically no difference (we shall come back to this point later
on).

\subsection{Discussion of results}

In Figs. 3 and 4 we show the reduced cross sections $\rho (p)$, for
proton knock--out from the shells $3s_{1/2}$ and $2d_{3/2}$ in
$^{208}$Pb, respectively, obtained in various approximations as
functions of the missing momentum $p$. The result of the
relativistic calculation (solid line) is compared in each figure
with the results of nonrelativistic calculations obtained with
the standard (long dashed line) and DEB (dashed line) optical
potentials. The dotted line shows the result obtained using the
solution of the Schroedinger equation with the DEB potential
multiplied by the factor $K(r)$.

One can clearly see in figures 3 and 4 that, taking the relativistic
result as a reference, the large discrepancy found in the
nonrelativistic calculations with the standard potential decreases
substantially when using the DEB potential. As expected, the
nonrelativistic result gets closer to the relativistic one when
using the optical potential that gives an equivalent fit to elastic
nucleon scattering. We recall that the standard potential is a
particular 15--parameter fit to 46 data on elastic proton scattering
at 98 MeV from $^{208}$Pb, while the relativistic potential (and
hence the DEB potential) is a global fit over a wide range of proton
energies and mass numbers involving more than 4000 data points.
These two fits are not equivalent and hence it is not surprising
that the two potentials --standard and DEB-- differ (see Fig. 1)
and produce different $(e,e'p)$ cross sections, the latter giving
more absorption for the cases studied here.

This allows us to conclude that a large part of the discrepancy
between relativistic and nonrelativistic results is reduced when
using relativistic and nonrelativistic optical potentials that
give equivalent fits to elastic proton--nucleus scattering. Yet,
it is also clear from Figs. 3 and 4 that even with the DEB potential
there are still sizeable differences between the nonrelativistic
result and the relativistic one. The latter is only recovered when
the Darwin factor $K(r)$ is also taken into account. This means
that the $(e,e'p)$ cross section is sensitive to the increased
reduction in the nuclear interior of the relativistic outgoing
nucleon density. This reduction is clearly seen when one plots
the ratio between the relativistic ($\bar{\Psi}_F \gamma ^0 \Psi_F$)
and the nonrelativistic ($\phi^\dagger \phi $) density profiles.
Said ratio is mainly given by the real part of $A(r)$ shown in Fig.
2. The effect of the Darwin factor is irrelevant to elastic
proton--nucleus scattering but is important in $(e,e'p)$ processes
that are sensitive to the nuclear interior.

One may wonder whether it is legitimate to use $K \phi$ when working
within the nonrelativistic formalism and whether the above comparison
is actually meaningful. It is easy to convince oneself that this is
indeed the case when using the {\sc dweepy} program as done here.
The simplest way to show that is to consider the direct Pauli
reduction \cite{Coop93,Fear94} of the relativistic nucleon current.
To carry out this reduction, four--component wave functions with
only positive energy components are built,
\begin{equation}
\Psi_+({\bf p})=N_+(p) \left( \begin{array}{c} \Psi_{\rm up}
({\bf p}) \\
\displaystyle \frac{\mbox{\boldmath $\sigma$}\cdot {\bf p}}{E(p)+M}
\Psi_{\rm up}({\bf p}) \end{array} \right)\; ,
\label{4compo+}
\end{equation}
out of the fully relativistic four--component wave functions using
the positive energy projection operator \cite{BD64}, and an
effective nonrelativistic current operator $\hat{J}^\mu_{\rm n.r.}$
is defined such that
\begin{equation}
\bar{\Psi}^F_+ \hat{J}^\mu_N \Psi^B_+=
\Psi^{F\, \dagger}_{\rm up} \hat{J}_{\rm n.r.}^\mu
\Psi^B_{\rm up}
\label{jnr}
\end{equation}
with $\Psi^B_{\rm up}$ and $\Psi^F_{\rm up}$ the upper components of
the relativistic initial and final wave functions. In practice, once
the operator $\hat{J}_{\rm n.r.}^\mu$ is obtained the nonrelativistic
current is calculated using initial and final wave functions that
are solutions of ordinary Schroedinger equations.

It can be shown \cite{Fear94} analytically that to third order in
$(p/M)$ the operator in equation (\ref{jnr})
${\hat J}^\mu_{\rm n.r.}$ is identical to the nonrelativistic
current operator obtained by the Foldy--Wouthuysen procedure and
used in the code {\sc dweepy}. We have also checked that up to
fourth order the differences are negligible (less than 0.1\% for
the energies considered here). Eq. (\ref{jnr}) is then a useful
bridge to understand the relationship between the relativistic
results and the nonrelativistic ones with the DEB potential in
Figs. 3 and 4, as well as its meaning.

To this end we first compare in Fig. 5  the fully relativistic
results (solid line) for the $3s_{1/2}$ and $2d_{3/2}$ shells in
$^{208}$Pb  with the results obtained with the relativistic code
when using initial and final nucleon wave functions projected on the
positive energy plane (dashed line).  One can see that the
differences between the results are small, being only noticeable
at relatively high $p$. This shows that the coupling to the negative
energy contributions of the Dirac solutions  does not play an
important role in the fully relativistic result and cannot be
responsible for the observed discrepancies between the relativistic
and nonrelativistic results  in Figs. 3 and 4.  It then follows,
taking also into account Eq. (\ref{jnr}) and the correspondence
between $\hat{J}_{\rm n.r.}^\mu$ and the current operator used in
{\sc dweepy}, that in order to recover the relativistic result one
has to use the upper components of the initial (bound) and final
(scattering) nucleon wave functions. This explains why in Figs.
3--4 the nonrelativistic calculations with the DEB potential
reproduce the relativistic result once the factor $K(r)$ is taken
into account.

Although not explicitly shown here, the situation is similar with
regard to the bound nucleon wave function, i.e., if in the
nonrelativistic calculation we use for the bound state the solution
($\phi_B$) of the Schroedinger equation with the DEB potential
corresponding to the relativistic S--V potential used in the
{\sc timora} code, we have to take into account an extra factor
$K_B (r)$. This reflects the fact that, as seen from Eqs.
(\ref{psiup})--(\ref{jnr}), the nonrelativistic current operator
consistent with the nonrelativistic solutions of DEB potentials,
is related to that in the right hand side of Eq. (\ref{jnr}) by
\begin{equation}
\hat{J}^{\rm eff}_{\rm n.r.} = K^*(r) \hat{J}_{\rm n.r.} K_B (r)
\label{jeffec}
\end{equation}
Since {\sc dweepy} uses  $\hat{J}_{\rm n.r.}$ rather than
$\hat{J}^{\rm eff}_{\rm n.r.}$, we cannot recover the relativistic
results when using $\phi({\bf r})$ and/or $\phi_B({\bf r})$ unless
we insert the corresponding $K$ factors. A similar remark was first
pointed out in Ref. \cite{sherif} in the context of photonuclear
reactions, where the effect of the S--V potentials in the interaction
Hamiltonian was studied up to second order in a $1/(E+M)$ expansion.

Clearly $\hat{J}^{\rm eff}_{\rm n.r.}$ depends on the relativistic
potentials used in the calculations and for the purpose of
comparing results obtained in the nonrelativistic framework, it is
advantageous to stick to a single definition of the current
operator, as that used by {\sc dweepy}, adding the required
modifications {\it a posteriori}. It should also be pointed out
that in standard nonrelativistic analyses, the bound nucleon wave
function fits observables  (rms radii, binding energies, etc.)
that depend on the nuclear interior. Thus, to the extent that this
nonrelativistic wave function fits similar phenomenology as the
fully relativistic one, the study of the effect of $K_B (r)$ is
not as meaningful, for the purpose of this paper, as that of
$K(r)$. This is why we focus here on $K(r)$.

It is interesting to compare the function $K(r)$ with the Perey
factor (PF) as defined in Ref. \cite{nonloc}:
\begin{equation}
f(r)=\exp\left( \frac{1}{2} \beta^2 \frac{M}{2\hbar^2}
V_{\rm C}\right) \; ,
\label{pf}
\end{equation}
where $\beta$ is a nonlocality range parameter and $V_{\rm C}$ is
the central potential in Eq. (\ref{vc}). Analogously to $K(r)$,
the PF produces also a reduction of the wave function in the
nuclear interior. In fact, the PF calculated with the DEB potential
has a similar shape to the function $K(r)$. It is worth pointing
out that the need for the PF emerged from a completely different
starting point. Namely, from the analysis of nonlocalities of the
nonrelativistic optical potential, parametrized in terms of the
nonlocality range parameter $\beta$.

In Fig. 6 we show the effect in the ($e,e'p$) reduced cross section
due to the inclusion of the PF $[f(r)]$. We show nonrelativistic
results obtained with the DEB potential and compare them with the
relativistic result. The results in this figure correspond to the
$2d_{3/2}$ orbital in $^{208}$Pb and to two values of the $\beta$
parameter ($\beta = 0.85,1.0$ fm). As can be seen in the figure the
effect of the PF is similar to the effect of $K(r)$ shown and
discussed in Figs. 3,4. Actually for $\beta=1$ the result obtained
with DEB+PF reproduces quite well the relativistic result. This is
consistent with the fact that for this $\beta$ value the depth
(0.83) of the PF is comparable to the depth (0.79) of $K(r)$.
With the most commonly used value of $\beta$ ($\beta=0.85$), the
effect of the PF is not sufficient to reproduce the relativistic
result.

We have checked that the effect of the PF for $\beta=1$ with the
DEB potential is similar to the effect of $K(r)$ also for the
$3s_{1/2}$ orbital in $^{208}$Pb and for the orbitals $2s_{1/2}$
and $1d_{3/2}$ in $^{40}$Ca also considered in Ref. \cite{Udietal}.
In table II we summarize the results for all these orbitals. In this
table we give, for each orbital and nucleus, the ratio between the
nonrelativistic and the relativistic reduced cross sections at
their maxima. Following the order of appearance in the table, the
five nonrelativistic cases considered are: 1) the standard optical
potential (given in Ref. \cite{Quint} for $^{208}$Pb and in Ref.
\cite{Kra89} for $^{40}$Ca), 2) the DEB optical potential, 3) DEB
including the Perey factor with $\beta=0.85$, 4) DEB including the
Perey factor with $\beta=1$, and 5) DEB including $K(r)$. From this
table it is clear that the trend observed in going from the
standard optical potentials to the DEB+$K(r)$ case is similar for
all the shells studied in Pb as well as in Ca. Compared to the
relativistic calculation, the standard optical potentials produce
too large ($e,e'p$) cross sections with ratios between 1.2 and 1.6.
These ratios are reduced to values between 1.1 and 1.3  when the DEB
potential is used. The nonlocal corrections introduced by the Perey
factor act in the right direction and the cross sections become
closer to the relativistic results, particularly for $\beta=1$,
where the agreement with the relativistic result is comparable to
that obtained with $K(r)$. With the inclusion of the function
$K(r)$ the relativistic results are recovered within a reasonable
1\% to 5\% deviation.

At this point one may wonder if the similarity between the effect
of $K(r)$ and PF for $\beta=1$ fm could be regarded as more than a
mere coincidence. Actually, also the function $K(r)$ is related to
nonlocalities. Indeed, as explained in Appendix A, the function
$K(r)$ appears when converting the Schroedinger--like equation
with a nonlocal potential for $\Psi_{\rm up}$ into an ordinary
Schroedinger equation with no first derivative terms. This agrees
with the generally accepted notion that the relativistic approach
may  already include an important amount of nonlocal effects in
the nonrelativistic formalism.  Nonlocal terms in the
nonrelativistic equation may be partly justified just as well as
the spin--orbit terms in the nonrelativistic equation is
justified by looking into the nonrelativistic limit of the Dirac
equation. It is well known that the central potential derived from
S--V potentials which are local and non energy dependent contains
a linear dependence on the energy. The relativistic optical
potentials themselves already contain some energy dependence,
although relatively weak for the energies of interest here
compared to the explicit energy dependence of the DEB potential
shown in Eq. (\ref{vc}).

However the analogy between the Perey and $K(r)$ factors has to be
considered with caution until a rigorous study of the role of
nonlocalities is made, starting from nonlocal analyses in both
relativistic and nonrelativistic formalisms. This goes beyond the
scope of this paper where we point out this numerical similarity
as a `striking coincidence' that may encourage further work
along these lines.

\section{Summary and final remarks}

In Ref. \cite{Udietal} we found that the relativistic optical
potentials from Ref. \cite{Hama} are able to explain simultaneously
the elastic proton--nucleus scattering data and the $(e,e'p)$ cross
sections, while the most commonly used nonrelativistic ones fail to
do that with reasonable spectroscopic factors. As in previous work
\cite{Udietal} the nonrelativistic calculations are done here with
the code {\sc dweepy} that uses as input local nonrelativistic
optical potentials and bound nucleon wave functions. In this paper
we investigate why the relativistic and the nonrelativistic optical
potentials lead to different ($e,e'p$) reduced cross sections, even
though both are fitted to elastic proton--nucleus scattering data.
To this end we have followed the already known procedure of
building a nonrelativistic optical potential from the relativistic
one that lead to the same elastic scattering observables.

By this procedure we obtain a nonrelativistic optical potential
(DEB), as well as a function $K(r)$  relating the upper component
of the Dirac solution with the solution of the Schroedinger
equation with the DEB potential. The function $K(r)$ is less than
one in the nuclear interior and goes to one asymptotically. It is
this latter fact that guarantees that the DEB potential fits
equally well as the relativistic one the elastic proton--nucleus
scattering data at proton energies of our concern here.

We find that the DEB potential differs from the standard
nonrelativistic potential and leads to lower ($e,e'p$) cross
sections. This reflects the fact that the two potentials
correspond to nonequivalent fits of elastic proton--nucleus
scattering. We find that the large discrepancy between the results
of relativistic and nonrelativistic calculations is partly reduced
when using the DEB optical potential, instead of the standard one,
in the nonrelativistic formalism. This shows that better agreement
between relativistic and nonrelativistic results is found when the
potentials used give equivalent fits to elastic proton--nucleus
scattering. Yet, even with the DEB potential the $(e,e'p)$ cross
section turns out to be larger than the corresponding relativistic
result. The latter is however  recovered in the nonrelativistic
calculation with DEB after  inclusion of the function $K(r)$,
showing the sensitivity of ($e,e'p$) to the behavior of the wave
functions in the nuclear interior.

The role of dynamically enhanced  lower components is not relevant
for the $(e,e'p)$ processes discussed here, as shown by the fact
that the relativistic calculations produce nearly the same results
independently on whether one uses the complete solutions of the
Dirac equations for initial and final nucleons or one uses their
positive energy projected counterparts. This is crucial to
understand why the results obtained with the nonrelativistic
formalism using the DEB potential and the Darwin factor $K(r)$
reproduce the results of the fully relativistic calculations
within at most a 1--5\% deviation.

The above mentioned results reproduce the fully relativistic ones
because they amount to a strict two--component reduction of the
nuclear current in which negative energy contributions, which are
small anyway in the $(e,e'p)$ processes, have been neglected. On
the other hand if one forgets about the Darwin factor $K$ and uses
the function $\phi$ corresponding to the DEB potential when
calculating the $(e,e'p)$ cross sections in the nonrelativistic
framework, one finds a sizeable deviation from the fully
relativistic result. This is in contrast to the case of elastic
proton scattering where the relativistic results are recovered
with the DEB potential independently on whether the factor $K$
is or not taken into account. Since the elastic
proton--nucleus scattering data are only sensitive to the
asymptotic behavior of the wave function, these experiments cannot
provide information on the function $K(r)$ in the nuclear interior.
Therefore, the behavior of this function, and its effects on
observables sensitive to the nuclear interior, are predictions of
the model.

We have compared the effect of the function $K(r)$ that has a
relativistic origin, with that produced by the Perey factor that
simulates the effect of nonlocalities in the nonrelativistic
optical potentials. We have shown that both functions produce
nearly identical absorption for a nonlocality parameter $\beta =1$
fm. Although at first sight it may look surprising that one arrives
to similar effects from quite different starting points, one should
keep in mind that also the function $K(r)$ can be related to
nonlocalities. Indeed, when one replaces Eq. (\ref{psiup}) into
Eq. (\ref{schr}), one ends up with a Schroedinger equation with a
nonlocal optical potential for the upper component of the Dirac
solution (see Appendix A). Thus, both the $K(r)$ and the Perey
factors can be interpreted as corresponding to contributions which
give effective local representations of nonlocal effects.

Independently of the interpretation of $K(r)$, what we may
undoubtedly conclude is that: {\it i)} If we compare the relativistic
density for the outgoing nucleon ($\bar{\Psi}_F\gamma ^0 \Psi_F$),
obtained with the S--V potential of Ref. \cite{Hama}, to the naively
defined nonrelativistic density $\phi^\dagger \phi$, obtained with
the corresponding DEB potential, the former shows a depletion in
the interior governed by $|K(r)|^2$. {\it ii)} This depletion plays
an important role in $(e,e'p)$ processes, and must be taken into
account in nonrelativistic calculations performed with the
{\sc dweepy} code by introducing the Darwin factors, which appear
in a proper nonrelativistic reduction of the nucleon current
operator. A similar modification of the interaction hamiltonian for
photonuclear reactions has been discussed in Ref. \cite{sherif}.

During the refereeing process of this manuscript a paper by Jin and
Onley \cite{lastjin} has appeared that also discusses comparisons
between relativistic and nonrelativistic  calculations for
$^{40}$Ca$(e,e'p)$$^{39}$K cross sections. The main conclusions of
these authors seem to agree with ours. In their case the relationship
between relativistic and nonrelativistic results and its comparison
to the effect of the Perey factor are somewhat different due to the
fact that they consider a different nonrelativistic scheme to the
one considered here.

\acknowledgements

One of us (JMU) is carrying out the work as a part of a Community
training project financed by the European Commission under Contract
ERBCHBICT 920185 and thanks H.P. Blok for useful comments. This work
has been partially supported by DGICYT (Spain) under Contract
92/0021-C02-01.

\appendix
\section{Derivation of the Schroedinger equivalent equation}
Starting from the Dirac equation (Eq. (\ref{dirac})) we write:
\begin{eqnarray}
A_- \Psi_{\rm up}-{\bf p}\cdot \mbox{\boldmath $\sigma$}
\Psi_{\rm down}&=& 0 \label{a1} \\
A_+ \Psi_{\rm down}-{\bf p}\cdot \mbox{\boldmath $\sigma$}
\Psi_{\rm up}&=& 0 \label{a2}
\end{eqnarray}
with
\begin{equation}
A_\pm=E-U_V-U_C\pm (M+U_S) \label{a3}
\end{equation}
applying ${\bf p}\cdot \mbox{\boldmath $\sigma$}$ to Eq. (\ref{a2})
one gets
\begin{eqnarray}
\mbox{\boldmath $\nabla$}^2\Psi_{\rm up}=-{\bf p}\cdot
\mbox{\boldmath $\sigma$} A_+\Psi_{\rm down}&=&
i \mbox{\boldmath$\sigma$}\cdot \hat{\bf r} \left( \frac{dA_+}
{dr}\right) \Psi_{\rm down} \nonumber \\
&& -A_+ {\bf p} \cdot \mbox{\boldmath $\sigma$} \Psi_{\rm down}
\label{a4}
\end{eqnarray}
using Eqs. (\ref{a1}) and (\ref{a2}) to eliminate $\Psi_{\rm down}$
from the second and first terms in the r.h.s. of Eq. (\ref{a4}),
respectively, and using the identity
\begin{equation}
i \mbox{\boldmath $\sigma$}\cdot {\bf \hat{r}}\;
\mbox{\boldmath $\sigma$} \cdot {\bf p} = \frac{d}{dr} - \frac{
\mbox{\boldmath $\sigma$} \cdot {\bf l}}{r} \; ,
\end{equation}
one gets:
\begin{equation}
\left[
\mbox{\boldmath $\nabla$}^2 +
\frac{1}{A_+}\frac{dA_+}{dr}\frac{\mbox{\boldmath$\sigma$}\cdot
{\bf l}}{r} + A_+A_-
- \frac{1}{A_+}\frac{dA_+}{dr}\frac{d}{dr}\right]\Psi_{\rm up}=0
\label{a5}
\end{equation}
This is an exact second order differential equation for
$\Psi_{\rm up}$ that can be interpreted as a Schroedinger--like
equation with a nonlocal potential. The Schroedinger equivalent
equation and the DEB potential defined in Eqs. (\ref{schr}) to
(\ref{vso}) is obtained after the elimination of the last term
(Darwin term) in Eq. (\ref{a5}). To this end one looks for a
transformation, $\Psi_{\rm up}({\bf r})=K(r)\phi({\bf r})$, under
which Eq. (\ref{a5}) transforms into an ordinary Schroedinger
equation ({\em i.e.}, with no first derivative terms) and such
that $ {\Psi_{\rm up}({\bf r})}_{r \rightarrow \infty}
\longrightarrow\phi({\bf r})$.
This determines $K(r)$ in Eq. (\ref{kr}).

\begin{figure}
\caption{Nonrelativistic central ($V_{\rm C}$) and spin--orbit
($V_{\rm so}$) optical potentials for $^{208}$Pb and 100 MeV
proton energy. Solid lines correspond to the potentials
[Eqs. (\protect\ref{vc})--(\protect\ref{vso})] obtained
after the reduction of the Dirac equation with the S--V relativistic
optical potentials of Ref. \protect\cite{Hama}. Dashed lines
correspond to the Schroedinger optical model with the parameters
of Ref. \protect\cite{Quint}.  }
\end{figure}
\begin{figure}
\caption{Real and imaginary parts of $A(r)$ in Eq.
(\protect\ref{ar}) using the S--V relativistic optical potentials
of Ref. \protect\cite{Hama} for $^{208}$Pb and 100 MeV proton
energy.}
\end{figure}
\begin{figure}
\caption{Comparison of various ($e,e'p$) reduced cross sections
for the shell $3s_{1/2}$ of $^{208}$Pb (see text).}
\end{figure}
\begin{figure}
\caption{Same as Fig. 3 for the shell $2d_{3/2}$.}
\end{figure}
\begin{figure}
\caption{Comparison of the fully relativistic results (solid lines)
with results obtained after projection of initial and final nucleon
wave functions on the positive energy plane (dashed lines).}
\end{figure}
\begin{figure}
\caption{Comparison of the relativistic ($e,e'p$) reduced cross
section to nonrelativistic ones obtained with the DEB potential
and Perey factors for two values of the nonlocality parameter
$\beta$ (see text).}
\end{figure}

\mediumtext

\begin{table}
\caption{Parameters of the standard nonrelativistic optical
potentials for $^{208}$Pb and $^{40}$Ca from Refs.
{\protect \cite{Quint}} and {\protect \cite{Kra89}},
respectively. Depths are in MeV and distances in fm.
\label{tabquint}}
\begin{tabular}{lccccc}
           & $V$   & $W$     & $W_S$ & $V_{SO}$ & $W_{SO}$ \\
$^{208}$Pb & 27.93   & 9.000 & 3.16 & 4.000 & -0.835 \\
$^{40}$Ca  & 26.97   & 7.177 & 0.   & 4.379 & -1.066 \\
\hline           & $r_V$   & $r_W$ & $r_S$ & $r_{SO}$ & $r_{WSO}$ \\
$^{208}$Pb & 1.223   & 1.137 & 1.272 & 1.116 & 1.088 \\
$^{40}$Ca  & 1.225   & 1.410 &       & 1.034 & 0.999 \\
\hline           & $a_V$   & $a_W$ & $a_S$ & $a_{SO}$ & $a_{WSO}$ \\
$^{208}$Pb & 0.719   & 0.742 & 0.622 & 0.676 & 0.719 \\
$^{40}$Ca  & 0.706   & 0.570 &       & 0.648 & 0.620 \\
\end{tabular}
\end{table}

\begin{table}
\caption{Ratio between various nonrelativistic and the fully
relativistic reduced cross sections for p--values close to the
maxima of the two outermost shells of $^{208}$Pb and $^{40}$Ca.
\label{tabtwo} }
\begin{tabular}{lcccccccc}
&\multicolumn{4}{c}{$^{208}$Pb}&\multicolumn{4}{c}{$^{40}$Ca} \\
&\multicolumn{2}{c}{$3s_{1/2}$}&\multicolumn{2}{c}{$2d_{3/2}$}&
\multicolumn{2}{c}{$2s_{1/2}$}&\multicolumn{2}{c}{$1d_{3/2}$} \\
p (MeV) & 0 & 190 & 100 & 180 & 0 & 150 & --140 & 110  \\
\tableline
Standard & 1.59 & 1.57 & 1.36 & 1.48 & 1.16 & 1.48 & 1.25 & 1.31 \\
DEB      & 1.11 & 1.30 & 1.14 & 1.29 & 1.11 & 1.32 & 1.09 & 1.19 \\
DEB nonlocal $(\beta=0.85)$ & 1.06 & 1.14 & 1.07 & 1.11 & 1.06 &
1.14 & 0.99 & 1.13 \\
DEB nonlocal $(\beta=1.0)$ & 1.00 & 1.01 & 1.00 & 1.01 & 1.03 &
1.07 & 0.97 & 1.07 \\
DEB K(r)     & 1.01 & 1.03 & 1.02 & 0.98 & 1.05 & 1.05 & 0.98 &
1.07 \\
\end{tabular}
\end{table}


\begin{references}
\bibitem{deW90}
P.K.A. de Witt Huberts, J. Phys. G {\bf 16}, 507 (1990); L.
Lapik\'as, Nucl. Phys. {\bf A553} 297c (1993).
\bibitem{Quint}
E.N.M.  Quint, Ph.D. thesis, University of Amsterdam (1988).
\bibitem{Kra89}
G.J. Kramer,  Phys. Lett. B{\bf 227} 199 (1989);
Ph.D. thesis, University of Amsterdam (1990).
\bibitem{GP}
C. Giusti and F. Pacati, Nucl. Phys. {\bf A473}, 717 (1987);
{\bf A485}, 461 (1988);
M. Traini, Phys. Lett. B {\bf 213},1 (1988).
\bibitem{GP4}
S. Boffi, C. Giusti, and F. Pacati, Nucl. Phys. {\bf 336}, 416
(1980); C. Giusti and F. Pacati, Nucl. Phys. {\bf A336}, 427 (1980).
\bibitem{Pick85} A. Picklesimer, J.W. Van Orden, and S.J. Wallace,
Phys. Rev. C {\bf 32}, 1312 (1985).
\bibitem{Pick87} A. Picklesimer and J.W. Van Orden, Phys. Rev.
C {\bf 35}, 266 (1987).
\bibitem{Pick89} A. Picklesimer and J.W. Van Orden, Phys. Rev.
C {\bf 40}, 290 (1989).
\bibitem{Joe} J.P. McDermott, Phys. Rev. Lett. {\bf 65}, 1991 (1990).
\bibitem{Jin}
Y. Jin, D.S. Onley, and L.E. Wright,  Phys. Rev. C {\bf 45}, 1311
(1992).
\bibitem{Udietal}
J.M. Ud\'\i as, P. Sarriguren, E. Moya de Guerra, E. Garrido, and
J.A. Caballero, Phys. Rev. C {\bf 48}, 2731 (1993).
\bibitem{theor}
V. R. Pandharipande, C. N. Papanicolas, and J. Wambach,  Phys. Rev.
Lett. {\bf 53}, 1133 (1984); Z.Y. Ma and J. Wambach, Phys. Lett.
B{\bf 256}, 1 (1991); C. Mahaux and R. Sartor, Adv. Nucl. Phys.
{\bf 20}, 1 (1991).
\bibitem{Wagner}
G.J. Wagner, Progr. Part. Nucl. Phys. {\bf 24} (1990).
\bibitem{Sher86}
H.S. Sherif, R.I. Sawafta, and E.D. Cooper, Nucl. Phys. {\bf A449}
709 (1986).
\bibitem{Boffi}
S. Boffi, C. Giusti, F.D. Pacati, and F. Cannata, Nuovo Cimento
{\bf 98}, 291 (1987).
\bibitem{Blok87}
H.P. Blok, L.R. Kouw, J.W.A. den Herder, L. Lapik\'as, and
P.K.A. de Witt Huberts, Phys. Lett. B {\bf 198}, 4 (1987).
\bibitem{Ray92}
L. Ray, G.W. Hoffmann, and W.R. Coker, Phys. Rep. {\bf 212}, 223
(1992); and references therein.
\bibitem{Hama}
S. Hama, B.C. Clark, E.D. Cooper, H.S. Sherif, and R.L. Mercer,
Phys. Rev. C {\bf 41}, 2737 (1990).
\bibitem{Coop93}
E.D. Cooper, S. Hama, B.C. Clark, and R.L. Mercer, Phys. Rev. C
{\bf 47}, 297 (1993).
\bibitem{Udi93}
J.M. Ud\'{\i}as,  Ph.D. thesis, Universidad Aut\'{o}noma de Madrid
(1993).
\bibitem{Fru84}
S. Frullani and J. Mougey, Adv. Nucl. Phys. {\bf 14}, 1 (1984).
\bibitem{BD64} J.D. Bj\"{o}rken and S.D. Drell, {\em Relativistic
Quantum Mechanics} (McGraw--Hill, N.Y., 1964).
\bibitem{Rose}
M.E. Rose, {\em Relativistic Electron Theory} (Wiley and Sons, 1961).
\bibitem{SW86}
B.D. Serot and J.D. Walecka, Adv. Nucl. Phys. {\bf 16}, 1 (1986).
\bibitem{HS}
C.J. Horowitz and B.D. Serot, Nucl. Phys. {\bf A368}, 503 (1981);
Phys. Lett. B {\bf 86},  146 (1979).
\bibitem{HSbook}
C.J. Horowitz, D.P. Murdock, and B.D. Serot, {\it Computational
Nuclear Physics}, K. Langanke, J.A. Maruhn, and S.E. Koonin eds.
(Springer--Verlag, Berlin, 1991).
\bibitem{Chin92} C.R. Chinn and A. Picklesimer, Nuovo Cimento
{\bf 105A}, 1149 (1992).
\bibitem{McV}
K.V. McVoy and L. van Hove, Phys. Rev. {\bf 125}, 1034 (1962).
\bibitem{Fear94}
H.W. Fearing, G.I. Poulis and S. Scherer, Nucl. Phys. {\bf  A570},
657 (1994).
\bibitem{sherif}
M. Hedayati--Poor and H.S. Sherif, Phys. Lett. B{\bf 326}, 9 (1994).
\bibitem{nonloc}
F. Perey and B. Buck, Nucl. Phys. {\bf 32}, 353 (1962);
H. Fiedeldey, Nucl. Phys. {\bf 77}, 149 (1966);
M.M. Giannini and G. Ricco, Ann. of Phys. (N.Y.) {\bf 102}, 458
(1976).
\bibitem{lastjin}
Y. Jin and D.S. Onley, Phys. Rev. C {\bf 50}, 377 (1994).

\end{references}
\end{document}